\documentstyle[preprint,prc,aps]{revtex}
\begin{document}
\preprint{RUB-MEP-7/95}
\draft
\title{Hadronic molecules: meson-baryon hybrids}
\author{M.Shmatikov
\footnote{Permanent address: Russian Research Center "Kurchatov
Institute", 123182 Moscow, Russia}}
\address{ Institut f\"ur Theoretische Physik\\
	  Ruhr Universit\"at Bochum, D-44780 Bochum, Germany\\
	  {\rm e-mail: michaels@hadron.tp2.ruhr-uni-bochum.de}}
\maketitle
\tighten
\begin{abstract}
The existence of hadronic molecular-type hybrids consisting of
a baryon and a meson is argued. Long-range interactions due to one-pion
exchange are shown to be strong enough to produce a loosely bound
state. Specific features of a molecular hybrid are discussed.
\end{abstract}
%
The advent of QCD made customary classification of
hadrons in terms of their quark content, ordinary baryons
and mesons being $qqq$ and ${\bar q}q$ states respectively.
Considerable efforts, pioneered by Jaffe \cite{Jaffe},
have been spent to explore the possible existence of more
complicated states containing four or more quarks. Investigations
in the framework of the naive quark model \cite{Weinstein,Dooley}
have shown that some mesonic states may have a predominantly
molecular structure in the form of loosely bound ${\bar K}K$
pairs. It should be noted that this conclusion was obtained
by means of an analysis of forces operating in the QCD realm
i.e. in terms of quark-gluon forces.
Quite different mechanism for generating multiquark deuteron-like
states was suggested by T\"ornqvist \cite{Tq1,Tq2}. Guided by
the fact that it is possible
to get the bound state of two nucleons by one-pion exchange only
(strictly speaking, the tensor part of the latter), he argued
existence of (loosely) bound states involving
either two vector mesons or a vector-meson and a pseudoscalar-meson.
Indeed, the results he obtained are rather encouraging since the one-pion
exchange mechanism was
shown to be strong enough to bind heavy ($D$-- and $B$--) mesons
\cite{Tq2}. However, for lighter mesons containing $K$, $K^*$
and vector $\rho$, $\omega$ mesons one-pion exchange
can provide about only half of the attraction required for
the emergence of a bound state.
Since binding energy is a result of an interplay
between an attracting potential and a kinetic energy of coupling
particles, a requirement to be
satisfied in the formation of a shallow bound state
for some simple-type potentials can be
formulated in terms of a scaling condition relating the strength
constant of the potential and the masses of the particles
\cite{Blatt}. Such a scaling condition was addressed in \cite{Ericson}
with emphasis on the Yukawa potential representing the central
part of the one-pion exchange interaction. Since the coupling constants
of the $\pi$-meson to hadrons is known, the scaling condition
allows to introduce the notion of a critical mass, i.e.
the minimal reduced mass of the components required for the
molecule formation \cite{Ericson}. The boundedness condition
is apparently satisfied for infinitely heavy mesons.
Manohar and Wise analyzed a system
involving two heavy and two light mesons (more precisely the
$QQ\bar{q}q$ system)
 to show that the one-pion exchange provides an
attraction strong enough to bind two mesons containing
$b$-quarks \cite{Manohar}.

All the molecular states considered in the papers cited above referred
to mesons (a system containing the vector $K^*$ meson
and a $\Sigma$- (or $\Xi$-) hyperon was analyzed in \cite{Ericson}
with the negative conclusion as to possible emergence of
a bound state). We consider a more 'exotic' system involving
a heavy meson and a nucleon.  If bound, it will manifest
itself as a baryon with the same flavor as the heavy quark.
Provided the binding energy is large enough, it will
be subject to weak decay only.

The choice of the components of a 'molecule-to-be' is dictated
by several requirements. The state-of-the-art understanding of
nonperturbative QCD mechanisms does not give any hope for
obtaining more or less reliable quantitative results. Thus
we are bound to consider a long-range force whose characteristics
are known (or can be extracted) from experiment, i.e.
the celebrated one-pion exchange. Pion-exchange being established
as a driving force, the choice of components becomes more
selective. First, the $\pi$-meson is known to couple to the
isospin 'charge' of a particle. Then the nucleon containing
3 light ($u$- and $d$-) quarks is singled out among other
(stable) baryons. Second, the $\pi$-meson does not couple
to a pseudoscalar particle, thus prompting the choice of a
vector meson as the second component. Finally, the large mass
of the latter is anticipated to ensure boundedness with even
weak enough pion-mediated forces. Note also that the sign
of the $\pi$-meson coupling to a (vector) meson depends on
its quark content: it will be shown below that attraction
occurs between the nucleon and the heavy vector antimeson
(i.e. (${\bar Q}q$) state).

The specific properties of the pion almost unambiguously determine
the quantum numbers of the meson-nucleon system. Indeed,
the pion-induced tensor force at nearly all distances, where
the one-pion-exchanged is distinguished, is known to be much
stronger than its central counterpart \cite{Weise}. At
the same time, the presence of non-zero orbital angular momenta
invokes additional repulsion due to the centrifugal barrier.
Thus we are urged to conclude that  the most favorable
situation occurs in the $S-D$ system, just as it happens
in the deuteron case. The vector-meson -- nucleon system may
have two values of the total spin equal to 1/2 and 3/2.
The former value seems to be more promising. Indeed, the
total spin $\vec{S}$ is the sum of spins of the components,
\begin{equation}
\vec{S} = \vec{\Sigma} + \frac{1}{2}\vec{\sigma},
\label{totspin}
\end{equation}
with the spin operators of the vector meson $\vec{\Sigma}$
and the nucleon $\vec{\sigma}$ being normalized as
$\vec{\Sigma}^2 = 2$ and $\vec{\sigma}^2 = 3$.
The central part of the one-pion-exchange potential is
known to be proportional to the scalar product of spins
of involved particles. Using (\ref{totspin}), it can
be cast in the form:
\begin{equation}
\vec{\Sigma}\vec{\sigma} = S(S + 1) -\frac{11}{4}
\label{sprod}
\end{equation}
One can infer readily from the expression above that the
spin scalar product under consideration is twice as large
for the doublet case ($S = 1/2$) than for the quartet one
($S = 3/2$). Finally, the strength of the
one-pion-exchange potential depends on the isospin of the
system. Heavy mesons (like nucleons) belong to
an isodoublet and the interaction in the isoscalar state is
3 times stronger than that for the
isovector state. Thus we conclude that the most favorable conditions
for obtaining a bound molecular-like state of the nucleon
and the vector meson take place in the state with the quantum
numbers
$J^{\pi}(T) = 1/2^{+}(0)$. In the $L-S$ basis this state
corresponds to the coupled ${}^2S_{1/2} - {}^4D_{1/2}$ waves and
as such, bears much similarity to the properties of the deuteron
itself.

We proceed now to the quantitative analysis of the considered
system and begin with the  parameter controlling emergence of a
zero-energy bound state for the Yukawa potential
 \cite{Blatt},
\begin{equation}
s = 0.5953\cdot 2\bar{m}\,\gamma\,\frac{V_0}{m^2_{\pi}}\,,
\label{spar}
\end{equation}
where $\bar{m}$ is the reduced mass of interacting particles,
$m_{\pi}$ is the $\pi$-meson mass and $V_0$ is the potential
strength 'unit' introduced in \cite{Tq2}:
\begin{equation}
V_0 = \frac{m^3_{\pi}}{12\pi}\,\frac{g^2}{f^2}\,.
\label{vz}
\end{equation}
 The ratio of coupling constants $g/f$ is determined
by the effective Lagrangian of (constituent) quark - pion
coupling
\begin{equation}
{\cal L}_{int} = \frac{g}{f}\,\bar{q}\gamma^{\mu}\gamma^{5}
\vec{\tau}q\,\partial_{\mu}\vec{\pi}\,,
\label{int}
\end{equation}
where $f$ is the pion decay constant and $g$ has the meaning
of an effective pseudovector quark-pion coupling constant.
At present we will not specify its numerical value but assume
only that it can be extracted either from the constant of the
$\pi NN$ coupling or from the width of the pionic decay
of the vector meson into a pseudoscalar one without serious
contradictions.
The parameter $\gamma$ in (\ref{spar}) combines spin-isospin
factors which depend on the quantum numbers of interacting
particles and of the system as a whole.

To make our conclusions more transparent, we compare
the $s$ parameter (\ref{spar}) for the (heavy) vector-meson
and nucleon system ($s_{VN}$) to its $NN$ counterpart:
\begin{equation}
\rho\equiv
\frac{s_{VN}}{s_{NN}} = \frac{\gamma_{VN}}{\gamma_{NN}}\cdot
\frac{2}{1 + m_N/m_V}\,.
\label{ratio}
\end{equation}
Each nucleon furnishes a 5/3 factor, and in the $NN$ system
one more factor 3 emerges in both ${}^1S_0$ and
${}^3S_1 - {}^3D_1$ states. At the same time, in the $VN$ system,
isospin and spin factors are equal (up to a sign) to 3 and 2
respectively (compare to (\ref{sprod})) yielding
\begin{equation}
\rho = \frac{2\cdot5/3\cdot 3}
{
(5/3)^2\cdot 3}\cdot\frac{2}{1 + m_N/m_V}\,.
\label{rati}
\end{equation}
If,
following the arguments of \cite{Ericson}, we consider the
$s_{NN}$ value as corresponding to the emergence of a bound
state with zero energy (strictly speaking, the near-threshold
virtual state in the ${}^1S_0$ channel), inspection of
(\ref{rati}) shows that, for given quantum numbers of the
$VN$ state, a bound state will emerge for the mass
of the vector meson $m_V \geq 5/7\,m_N$. This is a condition which is
readily satisfied for $B^*$ and even $D^*$ mesons.
The boundedness criterion, as formulated in \cite{Blatt},
\begin{equation}
s_{VN} \geq 1
\label{crit}
\end{equation}
proves to be more stringent. Indeed, substituting in
(\ref{rati}) the known $s_{NN}$ value
($s_{NN} \approx 0.33$), we arrive at the expression
\begin{equation}
s_{VN}\approx\frac{0.784}{1 + m_N/m_V}
\label{ss}
\end{equation}
showing that the condition (\ref{crit}) cannot be
satisfied even for an arbitrarily heavy vector meson. Note,
however, that the $s_{VN}$ parameter in the case of the
$B^*N$ system ($m_{B^*}\approx 5.3$~GeV) proves to
be $s_{VN}\approx 0.67$. This value which is twice as large
as that of the $NN$ counterpart. It can be concluded that
the conditions
for binding a heavy meson and a nucleon by the one-pion-exchange
forces are very favorable.

We proceed now to the quantitative investigation of the system
containing a heavy vector meson and the nucleon. The potential
of one-pion exchange operating between these particles
reads
\begin{equation}
V_{\pi}(r) = \frac{5}{3}\,V_0\,\kappa
\left[C_s\cdot\tilde{y}_0(m_{\pi}\cdot r)+
C_t\cdot\tilde{y}_2(m_{\pi}\cdot r)\right]\,.
\label{pot}
\end{equation}
The potential-strength parameter $V_0$ is defined in (\ref{vz}).
The factor $\kappa$ is the scalar product of the isospin Pauli
matrices equal to --3 and 1 in the case of the isoscalar and
isovector state respectively. The matrices $C_s$ and $C_t$ are
determined by the values of the spins and orbital momenta.
In the case of ${}^2S_{1/2} - {}^4D_{1/2}$ coupled channels
they read
\begin{equation}
C_s = \left(
\begin{array}{cc}
-2 & 0\\
0 & 1\\
\end{array}\right)\,;\:\:\:
C_t = \left(
\begin{array}{cc}
0& -1\\
-1& \sqrt{2}\\
\end{array}\right)
\label{matr}
\end{equation}
Finally, the $\tilde{y}_{0,2}$ functions in (\ref{pot}) are
the well known Yukawa-type functions
\begin{equation}
y_0(x) = \frac{\exp(-x)}{x};\:\:\: y_2(x) = \frac{\exp(-x)}{x}
\left(1 + \frac{3}{x} + \frac{3}{x^2}\right)
\end{equation}
regularized at small distances. We use the regularization employed
in \cite{Tq2}, which corresponds to introducing a monopole-type
formfactor at each vertex of $\pi$-meson coupling. The cut-off parameters
$\Lambda$
in both vertices are assumed to be the same.
Before proceeding to the results of computations it would be
relevant to stress the different role of heavy
vector mesons and antimesons. According to the classification scheme
of the Particle Data Group \cite{PDG} a heavy mesons $B$ has
the quark structure as follows: $B = (q\bar{b})$ where $q$ is
a light quark. The $\pi$-meson couples to a light quark only.
The vertices of the $\pi$-meson interaction with a particle and
an antiparticle differ  in sign due to negative $G$-parity of
pion. The strength of interaction of two particles with the
 isospin 1/2 is 3 times larger in the isoscalar case than in the
isovector one. The strongest attraction is therefore achieved
in the (isoscalar) $(N\bar{B}^*)$ system, while the one-pion
exchange in the $NB^*$ system provides attraction in the isovector
state only. This circumstance predetermines the flavor quantum
number of the hybrid state under consideration.

Using the potential (\ref{pot})
\footnote{Note that additional factor (--1) appears in the case of
$N\bar{B}^*$ coupling}
we solve the eigenvalue problem
to search for a bound-state solution of coupled Schr\"odinger equations.
As a check, we calculate the binding energy of the deuteron
and reproduce the experimental result
with good accuracy for the values of parameters
$V_0\approx 1.34$~MeV (which corresponds to the value of $\pi NN$
coupling constant $f^2_{\pi NN}/(4\pi)\approx 0.08$) and the
value of the cut-off parameter $\Lambda\approx 1.35$~GeV. Note,
that the $\Lambda$ value (which is actually under debate) exceeds by about
$10\%$  that used in computations of \cite{Tq2}. With this set of
parameters a bound $(N\bar{B}^*)$ state does not exist.
It appears with the
increase of the $\pi qq$ coupling constant
$g$ which results in a stronger attraction scaled by the
$V_0$ factor (\ref{vz}). With $g$ coupling constant increased
by $\approx 40\%$ the binding energy makes about 1~MeV. The
specific value of the binding energy is very sensitive to the
details of short-range interaction between components of
the molecule-type state. Beyond the one-pion-exchange mechanism
in the hierarchy of interaction
ranges we need to consider the two-pion exchange. A strong correlation
between
pions results in an attraction of shorter range, which
is usually simulated by the exchange of the scalar-isoscalar
$\sigma$-meson. The constant of its coupling to the
quark can be determined in the framework of
the linear $\sigma$-model. More precisely, in this model
this coupling constant is equal to that of quark pion-coupling, and
the latter can be related, owing to the equivalence of pseudoscalar and
pseudovector types of coupling, to the
$g$ coupling constant in the Lagrangian (\ref{int}).
Taking, rather arbitrarily, the mass of a constituent
quark involved in these relations equal to
$m_q\approx 1/3 \,m_N$ and the mass of the $\sigma$-meson
$m_{\sigma}\approx 2m_{q}$ (as prescribed by the
chiral-symmetry limit) we get an attractive potential
generated by $\sigma$-meson exchange. The addition
of such a potential, with the strength weakened for
exploratory purposes by a factor 0.1, deepens the
binding energy to $\approx 10$~MeV. The radial dependence
of the ${}^2S_{1/2}$ and ${}^4D_{1/2}$ components of the
wave function is exhibited in fig.1.
Characteristic distances $r_c$, corresponding to the binding
energy $E_b$, are of the order of
\mbox{$r_c\approx 1/\sqrt{2\bar{m}\,E_b}$},
where \mbox{$\bar{m}\approx 0.8$~GeV} is the reduced mass of the $NB^*$
pair. Hence for the case of \mbox{$E_b\approx 10$~MeV} we
get \mbox{$r_c\approx 1.5$~fm}.
One can see
that both components peak at $r\approx 0.5$~fm and fall down
rather slowly  to about 1.5~fm. Such behavior of the wave
functions
complies with the long-range character of the one-pion exchange
affected by strong attraction of shorter range.

Molecular-type hybrid state is to be expected as well
in the system including the nucleon and the $D^*$-meson.
However, the reduced mass of the $(ND^*)$ pair is
smaller than that of the $(NB^*)$ pair.
Components being more light, emergence of a near-threshold
bound states requires stronger attractive potential.
Corresponding increase of $g$ coupling constant with
respect to the $(NB^*)$ case makes $\approx 10\%$.

A qualitative analysis of the $(N\bar{B}^*)$ system with
zero isospin in combination with numerical calculations
indicates the existence of
a bound state with the $J^{\pi} = 1/2^+$ quantum numbers.
It has the properties of an isoscalar baryon with the
flavor of the $b$-(anti)quark and the mass
$m\approx m_N + m_{B^*}\approx 6.2$~GeV.
The exact value of the binding energy $E_b$ depends on the details
of short range interaction and cannot be calculated
without invoking (poorly known) QCD mechanisms operative
in the nonperturbative domain. It should be stressed
at the same time that the properties of the hybrid
$(N\bar{B}^*)$ molecular-type state depend crucially on
the $E_b$ value. Indeed, the vector $B^*$ meson, because
of small mass difference with its pseudoscalar
partner ($\Delta m = m_{B^*} - m_B\approx 46$~MeV \cite{PDG}),
decays through the emission of a $\gamma$-quantum. If the
binding energy is small, the hybrid state has a very
diffuse structure and it would be reasonable to
expect that the properties of the $\bar{B^*}$ meson
are not significantly affected by the presence of the nucleon.
In this case, the decay time of the hybrid molecular
state will be about the same as that of a free vector
meson.
For larger values of $E_b$ the decay will be hindered
by the diminished phase space volume and, finally,
for $E_b \geq \Delta m$ the hybrid state will be stable
with respect to both strong and electromagnetic decays.
In any case, its width is expected to be very small,
not exceeding the width of the $B^*$ vector meson.

\acknowledgements
The author is indebted to the members of the Mittelenergiephysik
Arbeitsgruppe and especially to Prof.Dr.M.F.Gari and
Dr.J.A.Eden for kind hospitality extended to him during
the stay at Bochum University.


\end{document}